\documentclass[11pt,twoside]{article}
\usepackage{./asp2014}

\aspSuppressVolSlug
\resetcounters

\begin{document}

\title{The Role of Opacities in Stellar Pulsation}
\author{S. M. Kanbur,$^1$ M. Marconi,$^2$ A. Bhardwaj,$^3$ R. Kundu,$^3$, and H. P. Singh,$^3$
\affil{$^1$ Department of Physics, SUNY Oswego, Oswego, NY 13126, USA; \email{shashi.kanbur@oswego.edu}}
\affil{$^2$ Osservatorio di Capodimonte, Napoli, Italy; \\\email{marcella.marconi@oacn.inaf.it}}
\affil{$^3$ Department of Physics \& Astrophysics, University of Delhi, Delhi, India; \email{anupam.bhardwajj@gmail.com}}}

% This section is for ADS Processing.  There must be one line per author.
\paperauthor{SMKanbur}{shashi.kanbur@oswego.edu}{ORCID_Or_Blank}{SUNY Oswego}{Department of Physics}{Oswego}{NY}{13126}{USA}
\paperauthor{MMarconi}{marcella.marconi@oacn.inaf.it}{ORCID_Or_Blank}{INAF}{Osservatorio di Capodimonte}{Napoli}{NA}{80131}{Italy}
\paperauthor{ABhardwaj}{anupam.bhardwaj@gmail.com}{ORCID_Or_Blank}{University of Delhi}{Department of Physics and Astrophysics}{Delhi}{Delhi}{110007}{India}

\begin{abstract}
We examine the role of opacities in stellar pulsation with reference to
Cepheids and RR Lyraes, and examine the effect of augmented opacities on the theoretical pulsation light curves in key temperature ranges. The temperature ranges are
provided by recent experimental and theoretical work that have suggested that the iron
opacities have been considerably underestimated. For Cepheids, we find that the augmented opacities have noticeable effects in certain period ranges (around $\log P \approx 1$)
even though there is a degeneracy with mixing length. We also find significant effects
in theoretical models of B-star pulsators.
\end{abstract}

\section{Introduction}
Opacity is a crucial quantity in theoretical stellar pulsation and evolution. It is an atomic dataset required in the calculation of the radiative flux, and hence, determines both the equilibrium structure of the star and its subsequent pulsation properties. Moreover, through the kappa and gamma effects, the opacity is an important source of pulsational driving and damping. 
The hydrogen and helium ionization regions in the outer envelope are such that the opacity increases upon compression, storing up energy during
compression and releasing it during the subsequent expansion, thus amplifying any departures from hydrostatic equilibrium \citep{cox1980}.

In order to obtain consistency between stellar evolution and pulsation, it is important for stellar pulsation models to reproduce observed pulsation characteristics at masses and luminosities mandated by stellar evolution calculations. This was the original motivation that led \citet{simon1982} to issue a plea for the recalculation of Rosseland mean opacities inasmuch as to reconcile observed and theoretical characteristics of bump and beat Cepheids. \citet{simon1982} found that this could be accomplished by augmenting the metal opacities to resolve a number of discrepancies between evolution and pulsation. This prompted two groups, the OP consortium \citep{Seaton1994} and the OPAL group \citep{iglesias1991} to recalculate Rosseland mean opacities using two different approaches. Both groups produced somewhat similar results and managed to resolve the bump and beat Cepheid discrepancies using the observed data at the time \citep{moskalik1992, kanbur1994}.

In recent years, there has been an explosion of accurate, multi-wavelength data for Cepheids and RR Lyraes due mainly to {transient} and/or microlensing surveys \citep[e.g.,][]{udalski1993, soszynski2008, soszynski2015}. However, there has also been recent experimental evidence that the iron opacities are still underestimated \citep{bailey2015}. The OP project were not able to treat iron with the full $R$-matrix technique due to computational limitations. Now \citet{pradhan2009} have started to recalculate iron opacities using this method; preliminary calculations suggest an increase in iron opacities translating to an increase in Rosseland mean opacities of about $15\%$ at high temperatures. In the present work we investigate the effect of increasing Rosseland mean opacities by a factor of 1.15 for the temperature range $5.1<\log T < 5.5$.

\section{Fourier Analysis}

\begin{figure*}[!t]
\begin{center}
\includegraphics[width=0.9\textwidth,keepaspectratio]{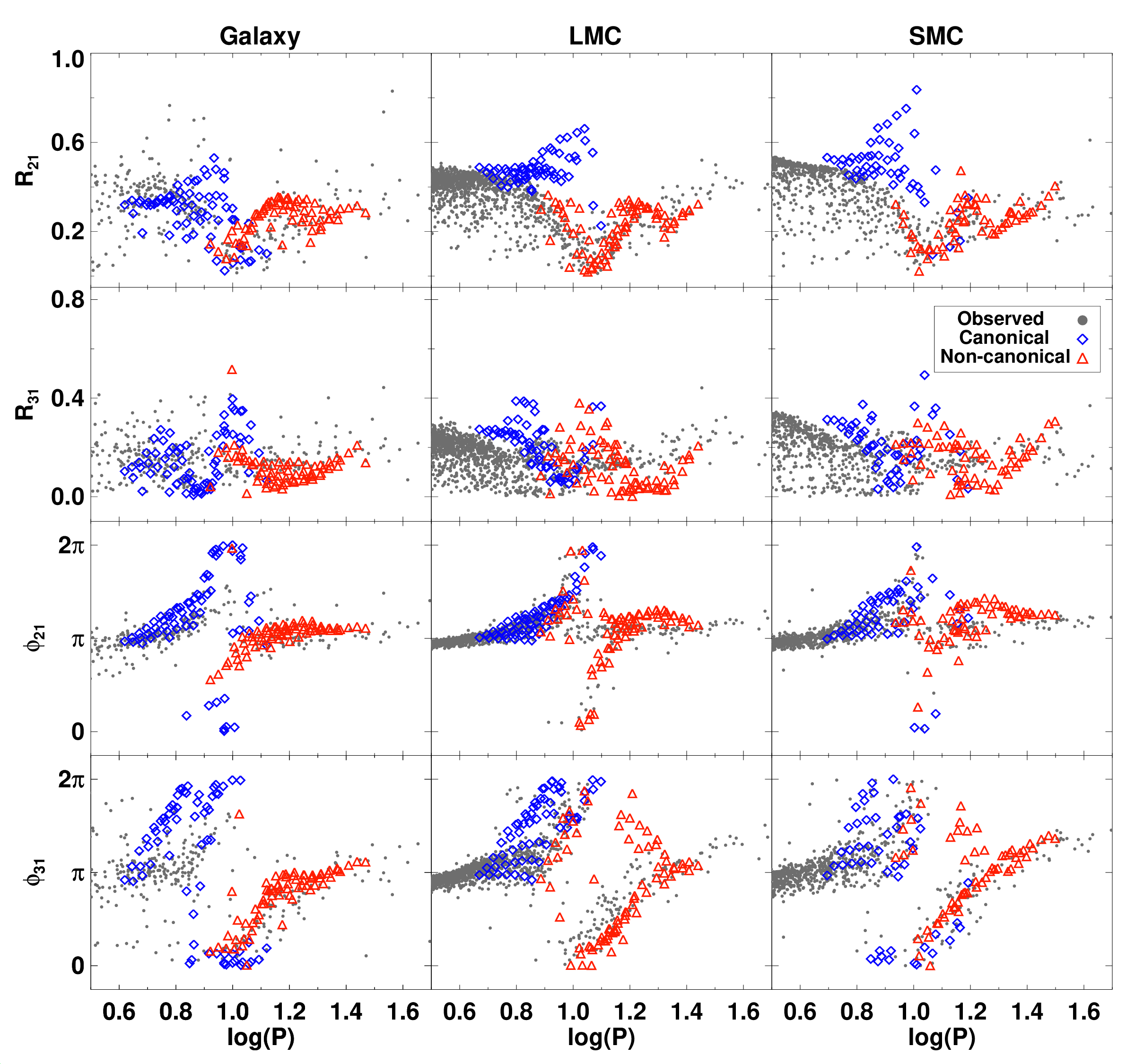}
\caption{Comparison of theoretical (red and blue open symbols) and observed
(black dots) Fourier parameters for Cepheid variables in the Galaxy and Magellanic
Clouds at I band. The theoretical models use canonical (blue open diamonds) and
non-canonical (red open triangles) ML relations.}
\label{fig:01}

\end{center}
\end{figure*}

\citet{slee1981} introduced Fourier analysis to quantify the non-linear structure of the Cepheid and RR Lyrae light curves, and applied this method to the data that were available then. In this method, both observational {\it and} theoretical light curves are subject to the following decomposition:
\begin{eqnarray*}
V & = & A_0 + \sum_{k=1}^{N}A_k\cos(k{\omega}t + {\phi}_k)\\
R_{k1} & = & A_k/A_1 \\
{\phi}_{k1} & = & {\phi}_k - k{\phi}_1\ ,
\end{eqnarray*}
where ${\omega}=2\pi/P$, the period $P$ being in days, and $N$ is the order of the fit. The preferred method of constraining theory with observations is then a comparison of the theoretical and observed Fourier parameters and their trend with the period. Fig.~\ref{fig:01} presents such results: we use Galactic and Magellanic Cloud data and recent full amplitude pulsation model light curves at a number of wavelengths \citep{bhardwaj2017}. The blue and red colors represent models made with canonical ($\log L =0.9+3.34\log M + 1.36\log Y - 0.34\log Z$) and non-canonical ($\log L$ is the canonical luminosity plus 0.25 dex) mass--luminosity (ML) relations. We note that the superb quality and quantity of observational data have provided a rigorous test-bed with which to constrain theoretical models. These results are discussed in detail in \citet{bhardwaj2015, bhardwaj2017}.

\begin{figure*}[!t]
\begin{center}
\includegraphics[width=0.9\textwidth,keepaspectratio]{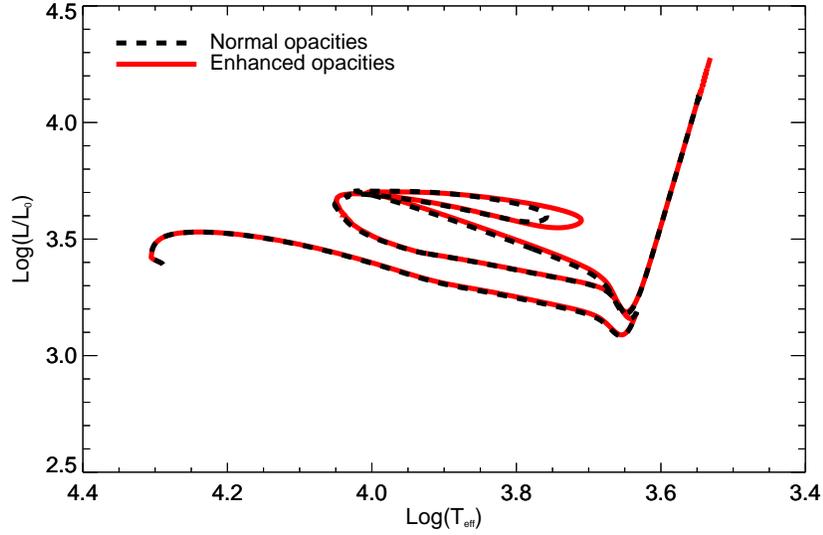}
\caption{Evolutionary tracks for a 7M$_\odot$  star with a $Z = 0.008; Y = 0.256$ composition. Solid red line: enhanced opacities. Dashed black line: normal opacities.}
\label{fig:02}
\vspace{-10pt}
\end{center}
\end{figure*}

\begin{figure*}[!t]
\begin{center}
\includegraphics[width=0.9\textwidth,keepaspectratio]{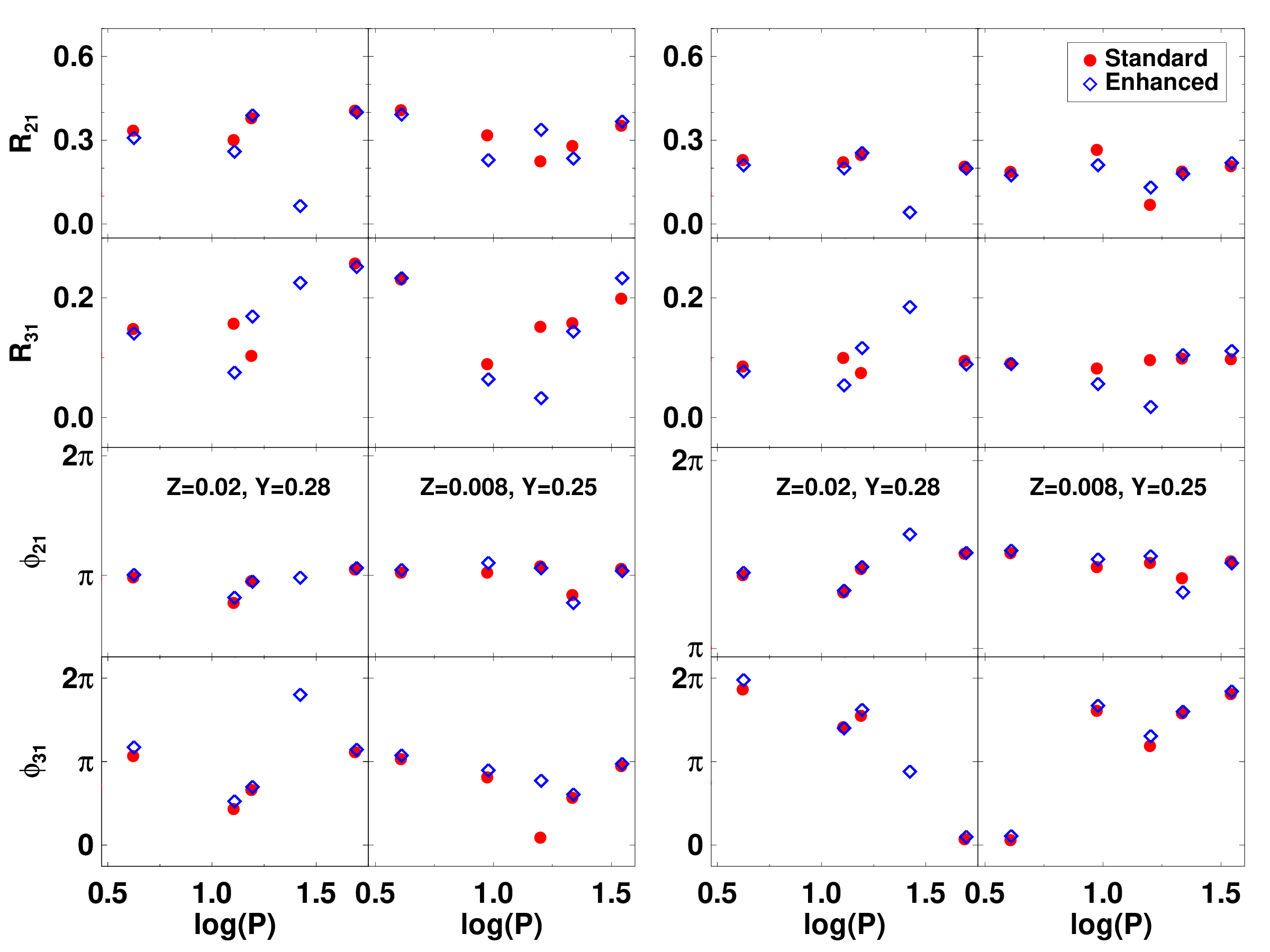}
\caption{Fourier parameters as a function of period with normal (solid symbols) and enhanced (open symbol) opacities. {\it Left panels:} V band. {\it Right panels:} K band.}
\label{fig:03}
\end{center}
\end{figure*}

\section{Augmented Opacities}

We computed a few pulsation models with augmented opacities (increased by a factor of 1.15 in the temperature range $5.1 < \log T < 5.5$). We also computed a sample of stellar evolutionary tracks with and without this artificial opacity enhancement using MESA. Fig.~\ref{fig:02} displays the results for a 7~M$_{\odot}$ star with a $Z=0.008; Y=0.256$ composition. The tracks with normal and enhanced opacity are quite similar except perhaps at the blue loop. We also computed a series of full amplitude Cepheid pulsation models spanning a range of periods with and without augmented opacities, $Z=0.02;Y=0.28$, with a canonical ML relation, and a fixed mixing length. We computed the theoretical light curves at $V/K$ wavelengths and carried out a Fourier analysis as described above. Fig.~\ref{fig:03} presents the Fourier parameters ($R_{k1}, {\phi}_{k1}$) plotted against the logarithm of the period (in days) at $V/K$ bands in the left/right panel, respectively. We see that the differences can be large at short wavelengths and in certain period ranges ($\log P\approx 1$). The differences in the Fourier parameters between the normal and enhanced opacity models are much larger than the errors in these parameters, though we note from Fig.~\ref{fig:01} and the results in \citet{bhardwaj2015, bhardwaj2017} that the changes in the mixing length parameters can also produce large differences in the Fourier parameters.

The superb quality of current observational data coupled with the Fourier method to study the nonlinear structure of Cepheid/RR Lyrae light curves do offer the possibility of using variable star observations to test newly computed opacities in stellar pulsation/evolution models.

%\articlefigure{fig02.eps}{fig:02}{Stellar Evolution for 7M$_\odot$ star with $Z = 0.008; Y = 0.256$ composition.}
%\articlefigure{fig3.eps}{fig:03}{Fourier parameters as a function of normal/enhanced opacities. The left/right panels show the V/K band respectively.}

\section{B Star Pulsations}

OP/OPAL/OPLIB opacities cannot reproduce the observed frequencies of B-star pulsators. \citet{dasznyska2002} studied the early main sequence B-type star {\it v Eri}, and modified the mean opacity profile to reproduce the observed range of frequencies. Their modifications were similar to the ones used in this paper. With this modification, they were able to reproduce observations. Here we tried a similar modification using MESA and GYRE with the following parameters: 7M$_{\odot}$,
$Y=0.2815$, and $Z=0.015$ considering only the fundamental and first overtone modes. Table~1 presents our results. In this table, we consider two mixing lengths, 1.5 and 5 and use both normal and enhanced opacity (as defined in the section on augmented opacities) models. We calculated the fundamental and first overtone period with GYRE whenever the track on the HR diagra, as computed by MESA, became pulsationally unstable. Since both fundamental (FU) and first overtone (FO) modes become unstable at a number of places, we have presented the period at all these points for this one $7M_{\odot}$ model. Our interest is primarily in the difference that enhanced opacities can make. We see that augmented opacities can indeed make a significant change in the theoretical periods and, hence, concur with previous work \citep{dasznyska2002}.

\begin{table}[!h]
\caption{LNA analysis of a $7M_{\odot}$ models appropriate for {\it v Eri} using
MESA+GYRE and normal and enhanced opacities.}
\smallskip
\begin{center}
{\small
\begin{tabular}{llllc}  % l = left, c = centered
\tableline
\noalign{\smallskip}
\tableline
\noalign{\smallskip}
Mixing Length 1.5\\
Period with normal opacities&&&&\\
FU&0.256774&7.51159&14.7145&17.5200\\
FO&&4.82029&9.45686&11.2258\\
Period with enhanced opacities&&&&\\
FU&0.260017&7.61011&15.2465&17.6854\\
FO&&4.80815&9.63397&11.5206\\
\hline
Mixing length 5\\
Period with normal opacities&&&&\\
FU&0.260284&5.94326&11.0175&12.9086\\
FO&&3.88381&8.16356&7.01949\\
Period with enhanced opacities&&&&\\
FU&0.259143&6.06253&11.3137&13.1524\\
FO&&3.81356&8.16799&7.09209\\
\noalign{\smallskip}
\tableline\
\end{tabular}}
\end{center}
{\footnotesize {\bf Notes:} Periods are in days. We use GYRE to compute the period every time the $7M_{\odot}$ becomes
unstable for the fundamental (FU) and first overtone (FO) models. Hence the period (in days) in
the last four columns correspond to different evolutionary phases while crossing the HR diagram.}
\end{table}

\acknowledgements SMK and HPS acknowledge support from IUSSTF provided to the Indo-US joint center on Theoretical Analysis of Variable Star Data in the Era of Large Surveys. RK thanks CSIR for a Junior Research Fellowship.

%%%%%%%%%%%%%%%%FIGURES%%%%%%%%%%%%%%%%%%%%%%%%%%%%%

% It is possible to reduce the size of a figure among other changes (see the instructions).  Here is an example:
%\articlefigure{fig01.eps}{fig:01}{Comparison of theoretical and observed Fourier parameters for Cepheid variables.}

%\articlefiguretwo{fig03a.eps}{fig03b.eps}{fig:03}{The left/right panel show the V/K band respectively.}

% It is possible to reduce the size of a figure among other changes (see the instructions).  Here is an example:

\clearpage % To force this stuff to happen by this point in the text, otherwise these will probably end up after the references.

\end{document}